# Design of PIC12F675 Microcontroller Based Data Acquisition System for Slowly Varying Signals

*N. Monoranjan Singh and K. C. Sarma.*
Department of Instrumentation & USIC, Gauhati University, Guwahati – 781 014, India.
*e-mail: mranjansn@yahoo.com

**Abstract:** The present paper describes the design of a cost effective, better resolution data acquisition system (DAS) which is compatible to most of the PC and laptops. A low cost DAS has been designed using PIC12F675 having 4-channel analog input with 10-bit resolution for the monitoring of slowly varying signals. The DAS so designed is interfaced to the serial port of the PC. Firmware is written in Basic using Oshonsoft PIC IDE and burn to the microcontroller by using PICkit2 programmer. An application program is also developed using Visual Basic 6 which allows to display the waveform of the signal(s) and simultaneously the data also can be saved into the hard disk of the computer for future use and analysis.

## 1. INTRODUCTION

A low cost PC based real time data logging system can be used in the laboratories for measurement, monitoring and storage of data for slowly varying signals in science and engineering stream[1]. Data logging and recording is a very common measurement application. In its most basic form, data logging is the measurement and recording of physical or electrical parameters over a period of time [2]. The frequency range of most of the bioelectric signals are slowly varying signals which range from 0.01 to 150 Hz but for EMG, it is from 5 to 2000 Hz [3]. Even though there are many biomedical instruments available, still research is going on to make a better, reliable and cost effective efficient instruments. Today many biomedical signal acquisition systems are PC based because of their high efficiency. There are many biomedical instrumentation systems available in the market for detection and analysis. These systems are used for training purpose and are very expensive [4]. NI DAQ card has a large number of functions which are not required for the present application and is also expensive. Thus, a custom novel cost effective hardware is designed, and the corresponding application software and firmware are also developed. For preventing missed data, polling technique is used that does not require a hardware interrupt. In Visual Basic, MSComm's OnComm event performs the event-driven routine which automatically jump to a routine when an event occurs. The application responds quickly and automatically to activity at the port, without having to waste time checking [5]. This application allows the custom control, display and storing of the recorded data to the PC.

## 2. MATERIALS AND METHODS
### 2.1 Data Acquisition Unit

A data acquisition system has been developed using PIC12F675 which is a mid range 8 Pin DIP microcontroller having four analog input channels and an internal crystal oscillator. Fig. 1 shows the circuit board of the DAS so designed and Fig.2 the circuit diagram. It consists of a 5 volt regulated power supply, PIC12F675 microcontroller and a MAX232 driver. PIC12F675 has four channels ADC with 10 bit resolution, in which the entire operation is controlled by the firmware. Pin 2 (GP5) of PIC12F675 is used to send the serial data to the PC through the MAX232 driver. The MAX232 converts the TTL signal from the microcontroller into RS232 voltage level. Since the PIC12F675 has an internal oscillator of 4MHz, so it works without any external clock. I/O pins GP0, GP1, GP2 and GP4 of IC2 are used as the analogue inputs. The MAX232 driver IC uses some external capacitors to enhance the voltage levels to RS232 level. A 9 Pin D Type female connector is used to connect to the COM port of the PC. The circuit is made into a PCB using ExpressPCB and the DAS is fabricated.

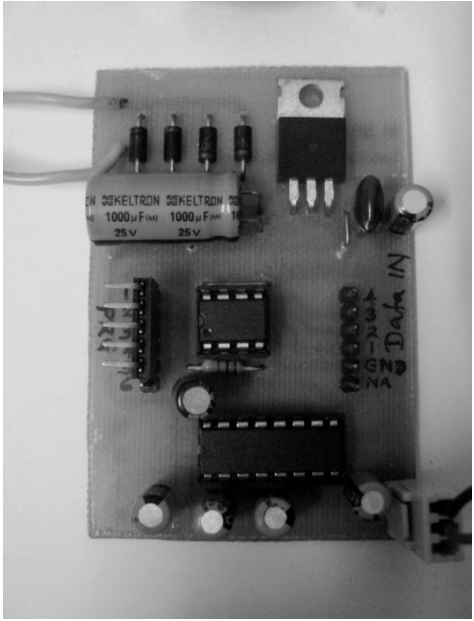

Fig. 1 Circuit Board for the DAS

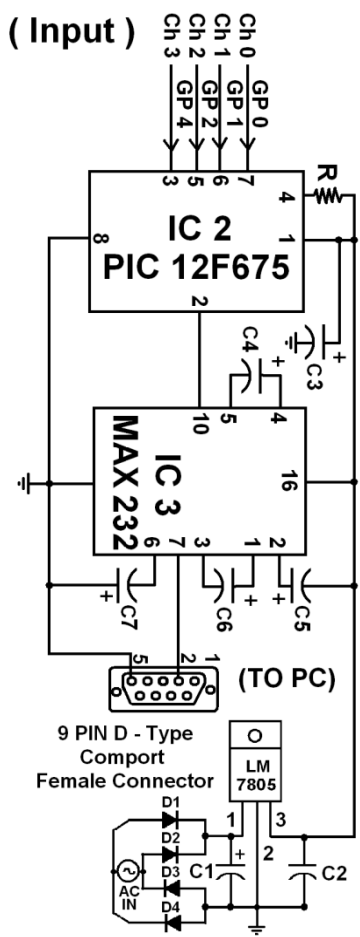

Fig. 2 Circuit diagram for the DAS

### 2.2 Signal source

A **GW Instek** Function Generator (Model No. SFG-1013) was used as a signal source and applied to the designed DAS in TTL, or after converting it to a uni polar signal (if it is in analog mode). Because, the DAS accepts voltage level between 0 to 5 volt for all the four analog inputs of the PIC microcontroller.

### 2.3 Software section

For the proper functioning of the data acquisition system, a firmware was developed and writes to the microcontroller and an application program is also developed in Visual Basic 6.

### 2.3.1 Firmware

A BASIC like program was written in Oshonsoft PIC Simulator IDE for proper ADC conversion at a fixed sampling rate and sending the digitized data serially. The program was compiled to make a hex file. The hex file so generated was written to the PIC12F675 microcontroller using PICkit2 programmer. The flowchart of the program so developed is given in Fig.3.

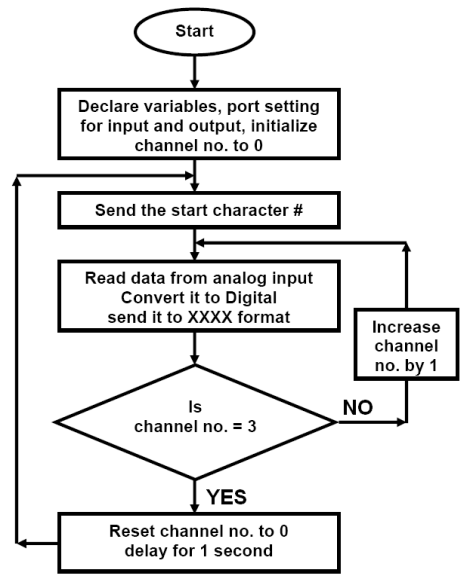

Fig. 3 Flow chart of the firmware

### 2.3.2 Application Program

For proper acquisition of the data by the PC, graphical display and saving into a file an application program is developed in Visual Studio 6.0. It uses the MSComm component to communicate through the comm port of the PC. The data so acquired is analysed and split into four values having four digit each, since it has 10 bit resolution it can read a value from 0 to 1023 for a channel. According to our channel selection a graph is plotted. All the data of the four channels is stored into the hard disk of the PC into .csv format. The data so stored can be used for future analysis and also the wave form can be reconstructed.

The on screen captured display of the application program is shown in Fig. 4.

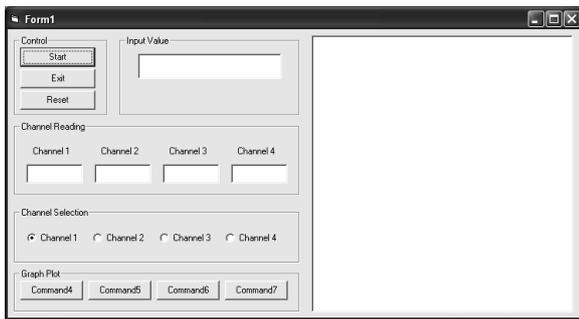

Fig. 4 Screen captured of the application program

## 3. RESULT AND DISCUSSION

DAS collects one sample per second. For studying the DAS, very low frequency signals are applied from the Function Generator. The acquired signals has been displayed numerically, graphically and the data is also stored into the hard disc of the computer in .csv (Comma Separated Value) format, which can be opened directly by MS Excel. The wave form can be reconstructed from the data of the stored file using MS Excel.

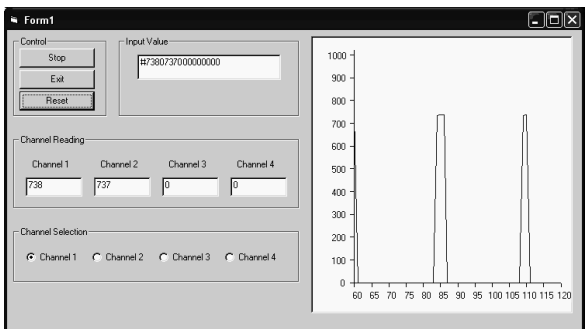

Fig. 5 Screen captured of the application program when input is given from Function Generator

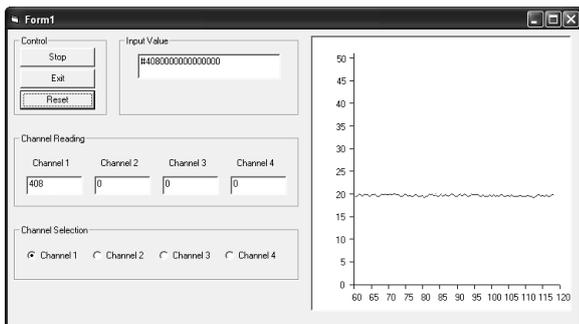

Fig. 6 Screen captured of the application program when input is given using temperature sensor LM35

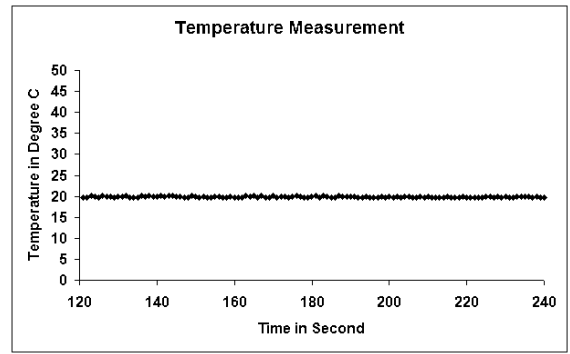

Fig. 7 Reconstructed waveform of the temperature using MS Excel from the stored file.

From fig. 6 and fig. 7, the above display and reconstructed graph are more or less same. It gives satisfactory result.

## 4. CONCLUSIONS

The designed DAS is a low cost with 10 bit resolution and compatible to PC and laptops, as serial port and USB are very common now a days. This DAS will be useful in Biomedical Instrumentation for the measurement of arterial pulse, ECG, EEG, etc. It will also be useful in research and practical laboratories where acquisition for slowly varying signals in science and engineering stream are required like temperature, humidity, light intensity etc. for the measurement, monitoring, analysis and storage of data. It can also be interfaced to the commonly available USB port of the PC or laptop using adapter BAFO BF-810.